# Physical requirements for scaling up network-based biocomputation

Jingyuan Zhu[1], Till Korten[2], Hillel Kugler[3], Falco van Delft[4], Alf Månsson[5], Danny Reuter[6,7], Stefan Diez[2,8], Heiner Linke[1,*]

1. NanoLund and Solid State Physics, Lund University, Box 118, 22100 Lund, Sweden
2. CUBE - Center for Molecular Bioengineering, Technische Universität Dresden, Dresden, 01307, Germany
3. Faculty of Engineering, Bar-Ilan University, Ramat Gan, Israel
4. Molecular Sense Ltd., Oxford, United Kingdom
5. Department of Chemistry and Biomedical Sciences, Linnaeus University, Kalmar, Sweden
6. Center for Microtechnologies, Technische Universität Chemnitz, Chemnitz, 09126, Germany
7. Fraunhofer ENAS, Technologie-Campus3, Chemnitz, 09126, Germany
8. Max Planck Institute of Molecular Cell Biology and Genetics, Dresden, 01307, Germany

E-mail: heiner.linke@ftf.lth.se

## Abstract

The high energy consumption of electronic data processors, together with physical challenges limiting their further improvement, has triggered intensive interest in alternative computation paradigms. Here we focus on network-based biocomputation (NBC), a massively parallel approach that benefits from the energy efficiency of biological agents, such as molecular motors or bacteria, and their availability in large numbers. We analyse and define the fundamental requirements that need to be fulfilled to scale up NBC computers to become a viable technology that can solve large NP-complete problem instances faster or with less energy consumption than electronic computers. Our work can serve as a guide for further efforts to contribute to elements of future NBC devices, and as the theoretical basis for a detailed NBC roadmap.

## 1. Introduction

The remarkable development of semiconductor technology, guided by the International Technology Roadmap for Semiconductors (ITRS) [1] and described by Moore's law [2] over the past five decades, has resulted in electronic processors that excel at fast and reliable processing of data in a sequential manner. However, significant further improvements in processor speed cannot be expected. Fundamentally, the processing of data at an increasing rate necessarily requires an increase in energy consumption per operation [3], because processing information in a shorter time is associated with higher dissipation [3-5]. Possibilities to reduce energy dissipation, for example by a further size reduction of electronic elements, are very limited due to physical size limits and the challenges of heat management. Therefore, there is wide agreement that Moore's law is coming to an end [6].

The limitations to processor operation rate also place a limit on the type of problems that can be solved. Specifically, many combinatorial problems fall into the NP or even NP-complete [7] categories, which require a computer to analyze a space of possible solutions that increases exponentially with problem size. For a sequentially operating computer, even the best algorithms currently known for NP-complete problems come at an exponential cost of energy and time. Examples of real-world applications that require the solution of such NP or NP-complete problems, and that are thus very costly to solve for electronic computers, include protein folding [8], planning in the logistics industry [9], model checking [10], as well as automated planning and scheduling in artificial intelligence [11, 12].

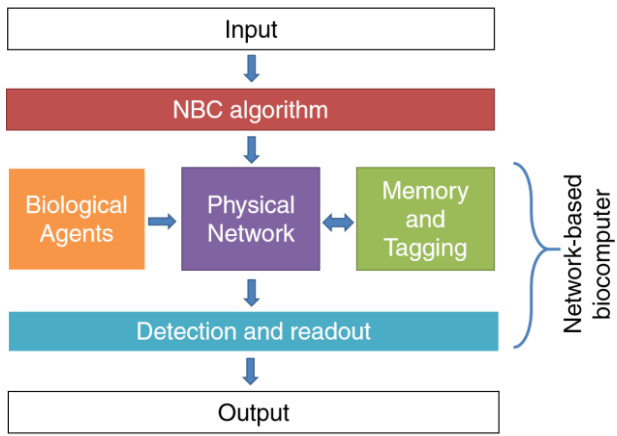

Figure 1 Architecture schematic of network-based biocomputing (NBC).

Alternative, parallel computing technologies, where a vast number of possible solutions to a problem are explored simultaneously, offer two fundamental advantages over sequential computers. First, a massively parallel approach can find the solution to a combinatorial problem faster, namely in polynomial rather than exponential time [13]. Second, parallel computers can fundamentally operate at appreciably lower energy cost per computation, regardless of problem size, because they can afford to keep the operation frequency low [3]. As a result, parallel computers are not subject to the same, fundamental trade-off between speed and energy use as electronic processors and may thus be able to solve much larger combinatorial problems with acceptable energy cost.

A number of alternative, parallel computing technologies are currently being explored. Out of these, DNA computing [14, 15] is highly energy-efficient [16, 17] and has been demonstrated to solve NP-complete problems [14, 18]. Whereas further upscaling appears currently limited by high error rates in DNA operating processes [19, 20], the technology holds promise for data storage [21]. Quantum computation [22, 23] offers unique potential for quantum simulations [24, 25], but upscaling remains highly challenging due to decoherence [26] and associated error rates [27, 28].

Here, we focus on another approach to parallel computation, namely network-based biocomputation (NBC) (**Figure 1**). In NBC a given combinatorial problem is encoded into a graphical, modular network that is embedded into a nanofabricated planar device. A large number of biological agents then explore all possible paths through the network in a parallel fashion. The exploration information fulfils the function of memory; this information is read and the output is the solution to the problem.

In a proof-of-principle experiment, NBC was used to solve, by brute force, a specific NP-complete problem (the subset-sum or SSP problem), using two types of molecular motors as the biological agents [3]. Owing to the high energy efficiency of molecular motors [29], already this first implementation was estimated to use several orders of magnitude less energy per operation than an electronic computer [3]. Alternative biological agents and their possible performance in NBC devices when scaling up to address NP-complete problems have been analysed [30].

Here, we analyse the fundamental requirements that need to be fulfilled to scale up NBC computers to become a viable technology that solves large NP problem instances faster or with less energy consumption than electronic computers. We define the four key physical elements of NBC (**Figure 1**) and discuss in detail the requirements on each element for scaling up NBC. In summary, the scale-up of NBC would require scalable agents (available in large quantity) with high moving speed and energy efficiency; a scalable physical network that ensures agent motion with negligible error rate; the ability to tag the agents in the network with negligible errors and lastly, the ability to detect single agents in parallel. Finally, we briefly discuss the contents of a future, detailed NBC roadmap.

## 2. Key elements of network-based biocomputation

The general principle of NBC is illustrated in **Figure 1**. As a first step, an algorithm is required that converts a given instance of a combinatorial problem into a graphical network. As an example we consider a generic combinatorial decision problem with a Boolean formula, F = f (A, B, C, …) where A, B, C… are Boolean variables (**Figure 2(a)**). An algorithm would then convert the problem instance into a directed graph that encodes the particulars of the Boolean formula. In general, the graph has entry nodes, exit nodes and many internodes (**Figure 2(a)**). The algorithm defines the local 'traffic rules' at each node that determine which path the arriving agent should take. Any path starting at an entry node, and ending at the exit node, encodes one unique assignment of the variables (e.g. A =1, B = 0, C = 1, see **Figure 2(a)**).

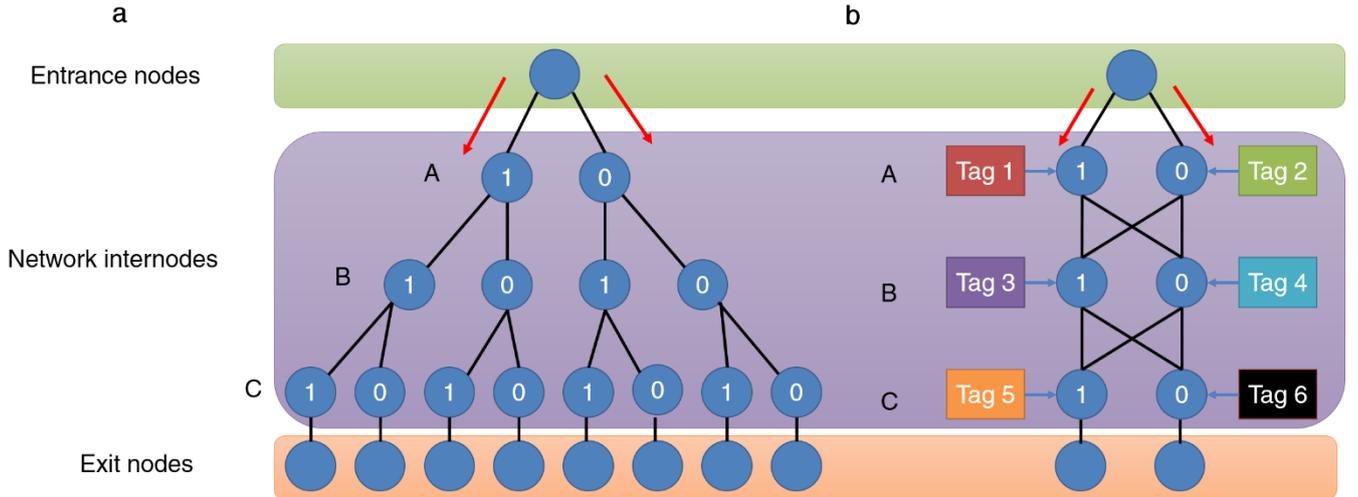

Figure 2 Schematics illustrating different encoding principles in NBC. a. Spatially encoded network. Each path corresponds to one possible assignment to the variables (A, B and C) and leaves the network at separate exits. b. Agent-encoded network. Here, still, each path corresponds to one possible assignment to the variables (A, B and C), but the different paths physically overlap at the exit, and tags are used to record the path information. Each agent visiting one of the two nodes representing any variable (A, B or C) receives a tag corresponding to the variable assignment (1 or 0). At the exit, agents need to be checked as to whether they collected all the target tags.

Crucially, all possible paths through the network then correspond to all possible assignments to the variables, namely, the problem's solution space. To find the correct solution, all possible paths through the network need to be stochastically explored by agents. For the combinatorial problems (see **Figure 2(a)**), each variable (2 possible assignments) would multiply the number of possible solutions by 2. Thus the number of possible solutions, *NoS,* grows exponentially with the number of variables, *NoV*.

$$NoS = 2^{NoV} \qquad (1)$$

Therefore, exponentially many agents are needed to solve the problem. Overall, the computation process within NBC includes the loading of agents, running of agents (exploration) and readout of the agents. The total computation time then is:

$$t_{all} = t_{loading} + t_{running} + t_{readout} \qquad (2)$$

Based on Figs. 1 and 2, we can now define key elements required for a scalable, physical implementation of NBC.

**Agents** are required in sufficiently high numbers to stochastically explore all possible paths through the network. The agents' size, speed and energy efficiency determine the required network size，obtainable computation speed and energy efficiency.

**Physical network.** Based on the NBC algorithm, a physical network with entry points and exit points needs to be fabricated. The network features channels that need to guide agents as required by the layout which is determined by the instance of problem (to be solved), for example ensuring unidirectional motion along paths and/or the ability to take turns or not to take turns at channel junctions.

**Memory and tagging.** In NBC, each agent may carry the information of one specific path through the network, and agents can in that case be considered as combined computing and memory units. In a spatially encoded network (**Figure 2 (a)**), each exit corresponds to a different solution. Readout happens by recording the position at which agents exit the network. More compact network designs are possible when information is encoded into the agents themselves, in the form of tags picked up by agents on the fly as they pass through specific network nodes (agent-encoded networks, see **Figure 2(b)**).

**Detection and readout.** Finally, the result of the computation must be read out by detecting agents and/or tags of agents at different exits of the network.

## 3. Requirements

In the following, we discuss each of the four key elements defined in Section 2 in more detail, with a focus on identifying the overall requirements of a scalable NBC technology.

### 3.1 Agents

**A1. Agents need to be available in large numbers and/or be able to multiply while they explore the network.** In NBC, each possible path through the network corresponds to one possible solution. Thus, the number of agents required, *NoA* is at least equal to *NoS*, and probably larger to ensure that all paths are explored stochastically (see also sections **N1** and **T1** below). If all agents need to be supplied through a limited number of entrances, the feed rate will necessarily limit the problem size that can be solved in a specific time frame.

Moreover, any serial feeding process of agents would cause $t_{loading}$, and thus $t_{all}$, to grow exponentially with *NoV*.

To overcome this limitation, one can consider multiplying the agents as they explore the network. Ideally, the number of agents (**NoA**) would double for each additional element in the problem, naturally adapting the processing power of the device to the number of combinations in the problem. *NoA* then becomes a function of time,

$$NoA_t = 2\, NoA_{t-\Delta t} \tag{3}$$

where $\Delta t$ is the time during which an agent explores one additional element (encoding one variable) in the network.

For instance, if single-celled bacteria are used as agents [31, 32], cell division could be used to increase the number of agents as they explore the network. In principle, both actin filaments and microtubules can be multiplied by splitting (using severing enzymes) followed by self-organized elongation by the addition of filament subunits (with and without the addition of filament stabilizing drugs or enzymatic filament polymerases) [33, 34].

**A2. Agents should have a small size and high running speed.** In NBC, a problem is solved when all paths through the network have been explored in a parallel manner. Regardless of the parallelism of the system, the time scale during which all agents explore the network is on the order of

$$t_{running} \propto \frac{L_p}{v} \tag{4}$$

where $v$ is the average running speed of the agent and $L_p$ is the length of a typical path through the network. Meanwhile, the $L_p$ is proportional to the product of length and width of the network:

$$L_p \propto L * W \tag{5}$$

Where L is the length of the network (entrance towards exit) and W is the width of the network.

In addition to the size of the encoded problem, the physical dimension of the network is also determined by the size of the agents. For example, a longer biological agent would require a longer path connecting the internodes so that each agent can be distinguished. We can assume that:

$$L_p \propto l \tag{6}$$

where $l$ is the characteristic dimension (length, diameter) of the agent.

Together, regardless of the network design, we can expect the following scaling:

$$t_{running} \propto \frac{l}{v} \tag{7}$$

Thus, to pursue a high computation power and to shorten the exploration time, the ideal agent should have a high speed and a small physical dimension $l$.

**A3. Agents need to move energy-efficiently.** The minimum energy cost of solving a problem by NBC is given by the energy required for all agents to propel themselves along all paths through the network[35]. For comparison with other computation techniques, it is useful to break this down to the energy cost of performing one computational operation, such as moving one agent from one internode to another.

One can estimate an energy consumption per operation of $2.5 \times 10^{-14}$ J/operation for the molecular-motor-based device demonstrated in [36] compared to about $3.6 \times 10^{-10}$ J/operation for the most advanced electronic computers [37], or an estimated minimum of $10^{-12}$ J/operation for microfluidics-based computers [38] (see [36] for a detailed comparison). The fundamental physical limit is provided by Landauer's principle [4], according to which a binary computational operation requires a heat generation of at least $k_B T \ln 2$ ($3 \times 10^{-21}$ J at room temperature).

## 3.2 Physical network

**N1. Negligible error rates at junctions.** An algorithm for solving a combinatorial problem by NBC essentially consists of a network architecture together with rules for how agents shall move through the network. A critical requirement is therefore that agents must not make errors such as taking wrong turns. As we will argue in the following, scalability requires a negligible error rate.

To understand how a finite error rate affects scalability, we consider that any agent that makes an error at least once may enter a path that produces an incorrect solution to the problem. Therefore, we can estimate the fraction of correct agents in the network as:

$$f_{correct} = (1 - p_e)^{NoN} \tag{8}$$

where $p_e$ is the probability for an agent to make a wrong turn at each node, and *NoN* is the number of nodes per path in the network. Similarly, we estimate the fraction of agents that make errors in the network as:

$$f_{incorrect} = 1 - (1 - p_e)^{NoN} \tag{9}$$

A detailed prediction of the distribution of the correct and incorrect agents depends on the precise network architecture. However, a preliminary requirement for the fraction of correct agents comes from the preferable positive signal to noise ratio in decibels (dB) which asks for:

$$f_{correct} \geq f_{incorrect} \tag{10}$$

By substituting equations ((9), (10), (11))we obtain the requirement of the probability of making errors:

$$p_e \leq 1 - \left(\frac{1}{2}\right)^{\frac{1}{NoN}} \tag{11}$$

By assuming large *NoN* and taking the first two terms in the binomial series of equation (11), we have:

$$p_e \leq \frac{5}{8 * NoN} - \frac{1}{8 * NoN^2} \tag{12}$$

Eq. 11 allows us to set an upper limit for the tolerable error rate for large *NoN*. For instance, a network with *NoN* = 50 requires that $p_e \leq 1.25\%$ while a larger network with *NoN* = 1000 would require that $p_e$ is smaller than 0.06%. We can use this to estimate the scalability limits of existing systems. For instance, the reported NBC with molecular motors [36] had $p_e$ = 2.1% at so-called pass junctions for the actin system, which thus would support a network with $NoN$ = 32. For the kinesin-microtubules systems, $p_e$ = 0.3% in [36], which would support a network with $NoN$ = 230 nodes. Any further increase in the achievable $NoN$ (scalability) requires a corresponding reduction in $p_e$.

A finite error rate also increases the $NoA$ required to solve the problem. Firstly, agents that make errors will become invalid and will not cover all the possible paths. To compensate for the loss, the required $NoA$ needs to be multiplied at least by the factor:

$$\alpha_{NoA,N} = \frac{1}{f_{correct}} = (1 - p_e)^{-NoN}, \quad (13)$$

that is, it increases exponentially with $NoN$ in the network unless $p_e$ is negligible (see **Figure 3**).

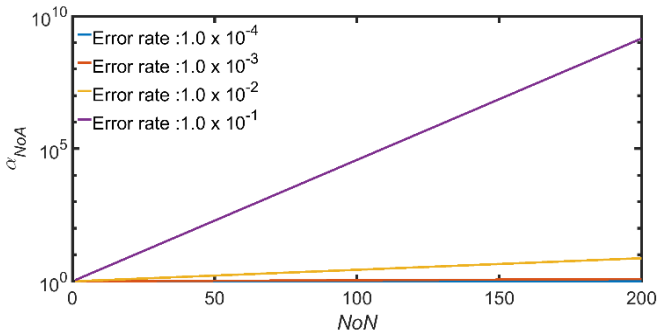

Figure 3 The *NoA* multiplication factor, which describes the increase in required *NoA,* as a function of *NoN* for different junction error rates (0.01%, 0.1%, 1% and 10%).

Second, agents that take wrong turns produce a background of incorrect solutions, further driving up the number of agents required to produce statistically reliable solutions. Thus, scalable NBC requires a negligible $p_e$. This can be achieved, for example, by using three-dimensional channel junctions [39], where filaments pass through tunnels and over overpasses to physically rule out wrong turns.

**N2. High-throughput nanofabrication technology.** Scalable NBC requires the ability to fabricate large-scale, planar devices with a very high spatial resolution to allow for small feature sizes for small agents (Requirement **A2**) and highly accurate junctions (Requirement **N1**), combined with high throughput to enable both, efficient optimisation and affordable production.

To initially demonstrate NBC with molecular motors, electron beam lithography (EBL) was used to fabricate networks [36]. However, because EBL is a serial process, the long time required to expose the resist while keeping a high resolution [40] would certainly become a bottleneck for the scale-up of NBC.

Parallel nanofabrication techniques could be considered such as extreme UV lithography (UVL) [41], or nanoimprint lithography [42, 43], which both enable a higher fabrication throughput of larger devices, while still maintaining a high resolution.

**N3. Programmable gating.** If the network layout hard-codes the problem instance, a new, unique physical network is required for each new problem instance. It would be highly advantageous to make the devices re-usable for solving different problem instances.

For the space-encoded network, each solution is represented by one specific path in the device. The ability to change pathways would enable programmable networks with dynamic encoding. Such programmable gating mechanisms could be created by switching the motility of filamentous agents in selective areas [44, 45]. It would enable a broad variety of permutations based on a dynamic algorithm, using just one standard large-scale network.

For the agent-encoded network, dynamic encoding may be achieved by modifying the tagging configurations in the path without altering the physical network. More detailed requirements for agent tagging will be discussed in Section 3.3.

**N4. Large bandwidth.** Agents must be provided with enough space to explore the network in parallel without crowding one another. We define the bandwidth of a network ($BW$) as the number of agents allowed to explore the network simultaneously. For a planar device, we define

$$BW \leq \frac{W}{w} \quad (14)$$

where $w$ is the width required for one agent (width of a single channel or width of a single agent) and $W$ is the width of the overall physical network.

The parallelism afforded by the network is limited by $BW$. Considering the fact that $NoA$ agents need to explore the network, and the time for each exploration is proportional to $l/v$ (equation (**7**)), one can then estimate the overall running time for all $NoA$ agents as:

$$t_{running} \propto \frac{NoA}{BW} * \frac{l}{v} \quad (15)$$

To speed up the computation and take advantage of the parallelism, a large BW is required which requires a physically larger network and benefits from smaller agents.

### 3.3 Memory and tagging

The requirement of tagging arises from the way in which information is stored in a network-based biocomputer.

For a space-encoded network as illustrated in **Figure 2(a)**, the spatial position of agents carries the exploration history and thus fulfils the role of memory. As $NoS$ grows exponentially for combinatorial problems, the spatial size as the memory of the network also needs to grow exponentially. For example, in **Figure 2(a)**, the network's width $W$ near the exits scales exponentially with *NoV*. The *L* would scale linearly with *NoV*. As a consequence, also the length $L_p$ of a typical path through the network scales exponentially:

$$L_p \propto L * W \propto NoV * 2^{NoV} \quad (16)$$

Such a fast-growing network would represent a bottleneck for the scale-up of NBC in terms of cost, fabrication time and operation time.

With physical computing agents such as cytoskeletal filaments or bacteria, information could also be stored on the agents themselves [30]. Such an agent-encoded network (**Figure 2(b)**) requires agents to be tagged on the fly as they pass through specific nodes in the network. Those locations correspond specifically to a certain assignment (e.g. A = true B = false and C = true). In this case, the memory function is assumed by the combination of tags encoded into the agent, while the agent's spatial position does not show the solutions but only the number of variables:

$$L_p \propto L * W \propto NoV \quad (17)$$

Therefore, the ability to tag the agents to encode their path information makes a network vastly more compact and scalable. The following requirements on tagging need to be met.

**T1. Tagging of agents with negligible error rate.** As the memory of agent-encoded NBC, the combination of tags on the agents carries information about possible solutions. During the operation, agents that miss some tags or pick up the incorrect tags would result in errors for computation. To exclude the result with wrong tags, error-correcting encoding can be introduced (see more details in **T2**). Meanwhile, agents that miss tags become invalid and, therefore, extra agents are required to explore the network to compensate for the loss of valid agents and find the correct solution. In the following analysis, we show that for problem instances with a large number of variables, the error rate in tagging needs to be negligible, because otherwise, the required *NoA* diverges.

To estimate the fraction of valid agents, we assume a tagging error rate as $p_{te}$ for all the variables. To compensate for the loss of valid agents, *NoA* needs to increase by a factor of $\alpha_{NoA,t}$.

$$f_{valid} = (1 - p_{te})^{NoV} \quad (18)$$

$$\alpha_{NoA,t} = \frac{1}{f_{valid}} = (1 - p_{te})^{-NoV} \quad (19)$$

As an example, if we ask for $\alpha_{NoA,t}$< 10, a network with NoV =100 variables would require $p_{te}$ < 2.3%. The scale-up of NBC (increase of the number of variables) requires a negligible $p_{te}$.

**T2. The number of possible combination of tags must be sufficiently large.** The size of the memory must ensure that each possible solution can be represented by a unique combination of tags. We require the number of unique tags:

$$NoT \propto NoV \quad (20)$$

Furthermore, there could be errors during the tagging. To identify errors one can introduce correcting codes, which then require the encoding of additional bits.

### 3.4 Detection and readout

To read out the information from the 'memory', single-agent detection is required to give information about the spatial position of the agents and/or the tagging information encoded into the agents. To maintain the advantage of parallel computing, detection should be done at a frequency on the order of $BW * v/l$ (see equation (**15**)) and at all network exits in parallel.

**D1. Spatial position reading in parallel.** For a purely space-encoded network, it is sufficient to count agents as they leave each exit that needs to be monitored in order to determine a solution. In the existing proof-of-principle experiments, this is done by fluorescence microscopy, using fluorescently marked filaments [36]. However, this approach is limited by the relatively small field of view at the high magnification needed to resolve very small agents (a typical field of view is only about 0.35 mm with field number ( diaphragm size of the eyepiece in mm unit) 28 and an 80x objective) and is by itself computationally intensive.

An alternative would be electrical readout, for example using nano field-effect transistor sensors based on single nanotubes or nanowires at each exit of interest [46]. Such detectors can be monitored in parallel, in real-time, and the electric signal offers the additional advantage of compatibility with electronic computers.

**D2. Tag reading in parallel.** For agent-encoded NBC, the agents carry their exploration history on themselves via the collection of tags along the route. For agent-encoded networks, it is necessary to read out the tags on each of the *NoA* agents, ideally in parallel. This is a challenging task, and the use of any tagging technology must be chosen with both the encoding mechanism and the readout mechanism in mind.

### 4. Comparison to electronic computers

To put the analysis above into context, an interesting benchmark is a direct comparison between NBC and an electronic computer.

An electronic computer operates at clock frequencies on the order of GHz [47]. In comparison, network-based computation operates at clock frequencies on the order of Hz [36]. Therefore, on the order of a billion agents are needed to be able to compete with electronic computers in problem solving capacity. This corresponds to only tens of nanograms of protein for cytoskeletal filaments agents or a few milligrams of bacteria (E. coli). However, providing a parallel channel for each agent is not feasible: it would result in network sizes on the order of metres for cytoskeletal filaments and kilometres for bacteria. This limit requires that either the same channels have to be used by multiple agents [36] - reducing parallelism or better, one should use efficient algorithms that discard redundant paths [48].

**Table 1** Summary of key elements and relevant parameters

| | Parameters | Relationship | Equations | Description |
|---|---|---|---|---|
| **Problem instance** | $NoV$ | | | Number of variables (problem instance size) |
| | $NoS$ | $2^{NoV}$ | (1) | Numer of possible solutions (solution space size) |
| **Agents** | $NoA$ | $> 2^{NoV}$ | (3)(13)(19) | Number of agents required |
| | $v$ | | | Velocity of agents |
| | $l$ | | | Length of agents |
| | $w$ | | | Width of agents/ width of a single channel |
| **Physical Network** | $L$ | $\propto NoV * l$ | (4) | Length of the network |
| | $W$ | $\propto 2^{NoV} * l$ | (16) | Width of the network (space-encoded) |
| | | $\propto l$ | | Width of the network (agent-encoded) |
| | $L_p$ | $\propto NoV * 2^{NoV} * l$ | (4)(6)(16) | Typical length of one agent path (space-encoded) |
| | | $\propto NoV * l$ | (4)(6)(17) | Typical length of one agent path (agent-encoded) |
| | $BW$ | $\dfrac{W}{w}$ | (14) | Bandwidth of the network |
| | $p_e$ | | (12) | Error rate per junction structure |
| | $NoN$ | | | Number of nodes per path in the network |
| | $f_c$ | $(1 - p_e)^{SoN/l}$ | (8) | Fraction of agents which didn't make any errors |
| **Memory and Tagging** | $p_{te}$ | | | Error rate per tagging event |
| | $f_{valid}$ | $(1 - p_{te})^{SoN/l}$ | (18) | Fraction of valid agents which are tagged correctly |
| | $NoT$ | | (20) | Number of unique tags |
| **Detection and Readout** | $t_{readout}$ | | | The time needed for readout |
| **Performance** | $t_{running}$ | $\propto \dfrac{NoA}{BW} * \dfrac{l}{v}$ | (4)(7)(15) | Total agents running time |
| | $t_{loading}$ | | | Agent loading time |
| | $t_{all}$ | $t_{loading} + t_{running} + t_{readout}$ | (2) | Overall computing time for one problem instance |

In addition, in order to encode billions of solutions, tags need to store more than 30 bits of information and billions of tags need to be analysed in parallel. The most feasible way to achieve this currently appears to be the use of DNA tags in combination with molecular biology methods such as Polymerase chain reaction (PCR) and next-generation sequencing for readout.

## 5. Discussion and conclusion

We introduced the four physical elements (agents, physical network, memory and tagging, readout and detection) required for NBC and identified the key parameters that describe their performance (Table 1).

In addition to self-propelled biological agents, such as bacteria and motor-driven filaments considered above, network-based parallel computing can in principle also be realized with any other type of agent. In evaluating candidate agents, the requirements defined above can be used.

For example, photons have been used as agents in a proof-of-principle experiment for optical, network-based parallel computing [48]. The use of photons as agents has two key advantages: first, they can be easily generated in large numbers, which can be sufficiently used to probe all the paths in parallel. Second, the photons move at the speed of light making the $t_{running}$ very short, even for a large and complicated network. However, in existing experiments, the photonic elements required to guide the light are rather large (30 mm between internodes and 690 mm for the network encoding SSP (2, 5, 7, 9) [48]). Based on the equation (**6**) and (**16**), the physical size of the device is estimated as:

$$W \propto 30 * 2^{NoV} \,(mm) \qquad (21)$$

Thus the physical size of a device with large $NoV$ would very quickly become unmanageable. For comparison, the network encoding the same problem in the molecular motor (actin-myosin) system has a size of 8 μm between internodes and, only 0.184 mm for the network [36]. In addition, the options to tag the photons are limited to polarization, with only two possible states.

Another interesting option would be to use synthetic molecular motors such as DNA walkers [49] or DNA navigators [50] in which the DNA strands are guided along the designed path and perform the computation. This walking behaviour benefits from the advantage of the sequence-specific recognition ability of DNA and solid-phase substrates. The approach is appealing in several aspects. First, the process of DNA recognition is a reversible process that is highly energy efficient. Second, as a structural material in its own right, DNA strands can be easily and precisely tagged. With the four bases of nucleic acid, DNA has a high storage capacity. Third, DNA molecules could be designed and replicated via biochemical methods. Fourth, the DNA strands are small in dimensions and can be confined by a nanofabricated track, with advantages for scalability. Challenges that still need to be solved include the development of faster walkers and the ability to track individual paths of many agents in parallel by single-molecule characterization.

## 6. Outlook

Progress in NBC requires a truly interdisciplinary approach combining mathematics, computer science, molecular biology, nanofabrication and scale-up and integration. These efforts include the development of new algorithms and the design of efficient networks; the discovery of new candidate biological agents with prefered speed and size; the reduction of errors by architectural improvements; the need for negligible error rate tagging method and effective, parallel read-out. Additionally, and beyond the scope of the present fundamental analysis, scalable methods for the system integration of all these technologies will be required.

Our analysis shows that many requirements are interdependent, and the performance of one aspect (e.g. the junction error $p_e$) has consequences for others (e.g. the *NoA* that needs to be supplied and detected). To more completely evaluate the scalability of a given NBC technology, therefore, more work is needed towards a 'figure-of-merit' for the performance of network-based biocomputers based on the discussion above in a roadmap for NBC.

## Acknowledgements

We thank Thomas Blaudeck (Fraunhofer ENAS) for valuable comments. This work has received funding from the European Union's Horizon 2020 research and innovation programme under Grant agreement no. 732482 (Bio4Comp).